\documentstyle[twoside,psfig]{article}

\catcode`\@=11
\long\def\@makefntext#1{
\protect\noindent \hbox to 3.2pt {\hskip-.9pt  
$^{{\eightrm\@thefnmark}}$\hfil}#1\hfill}

\def\@makefnmark{\hbox to 0pt{$^{\@thefnmark}$\hss}}	
	
\def\ps@myheadings{\let\@mkboth\@gobbletwo
\def\@oddhead{\hbox{}
\rightmark\hfil\eightrm\thepage}   
\def\@oddfoot{}\def\@evenhead{\eightrm\thepage\hfil
\leftmark\hbox{}}\def\@evenfoot{}
\def\sectionmark##1{}\def\subsectionmark##1{}}

\oddsidemargin=\evensidemargin
\newcounter{sectionc}\newcounter{subsectionc}\newcounter{subsubsectionc}
\renewcommand{\section}[1] {\vspace{12pt}\addtocounter{sectionc}{1} 
\setcounter{subsectionc}{0}\setcounter{subsubsectionc}{0}\noindent 
	{\tenbf\thesectionc. #1}\par\vspace{5pt}}
\renewcommand{\subsection}[1] {\vspace{12pt}\addtocounter{subsectionc}{1} 
	\setcounter{subsubsectionc}{0}\noindent 
	{\bf\thesectionc.\thesubsectionc. {\kern1pt \bfit #1}}\par\vspace{5pt}}
\renewcommand{\subsubsection}[1] {\vspace{12pt}\addtocounter{subsubsectionc}{1}
	\noindent{\tenrm\thesectionc.\thesubsectionc.\thesubsubsectionc.
	{\kern1pt \tenit #1}}\par\vspace{5pt}}
\newcommand{\nonumsection}[1] {\vspace{12pt}\noindent{\tenbf #1}
	\par\vspace{5pt}}

\newcommand{\textlineskip}{\baselineskip=13pt}
\newcommand{\smalllineskip}{\baselineskip=10pt}

\def\eightcirc{
\begin{picture}(0,0)
\put(4.4,1.8){\circle{6.5}}
\end{picture}}
\def\eightcopyright{\eightcirc\kern2.7pt\hbox{\eightrm c}} 

\newcommand{\copyrightheading}[1]
	{\vspace*{-2.5cm}\smalllineskip{\flushleft
	{\footnotesize ~}\\
	{\footnotesize ~}\\
	 }}

\def\abstracts#1#2#3{{
	\centering{\begin{minipage}{4.5in}\footnotesize\baselineskip=10pt
	\parindent=0pt #1\par 
	\parindent=15pt #2\par
	\parindent=15pt #3
	\end{minipage}}\par}} 

\renewenvironment{thebibliography}[1]
	{\frenchspacing
	 \ninerm\baselineskip=11pt
	 \begin{list}{\arabic{enumi}.}
        {\usecounter{enumi}\setlength{\parsep}{0pt}     
	 \setlength{\leftmargin 12.7pt}{\rightmargin 0pt} 
         \setlength{\itemsep}{0pt} \settowidth
	{\labelwidth}{#1.}\sloppy}}{\end{list}}

\newcounter{itemlistc}
\newcounter{romanlistc}
\newcounter{alphlistc}
\newcounter{arabiclistc}

\newcommand{\fcaption}[1]{
        \refstepcounter{figure}
        \setbox\@tempboxa = \hbox{\footnotesize Fig.~\thefigure. #1}
        \ifdim \wd\@tempboxa > 5in
           {\begin{center}
        \parbox{5in}{\footnotesize\smalllineskip Fig.~\thefigure. #1}
            \end{center}}
        \else
             {\begin{center}
             {\footnotesize Fig.~\thefigure. #1}
              \end{center}}
        \fi}

\newcommand{\tcaption}[1]{
        \refstepcounter{table}
        \setbox\@tempboxa = \hbox{\footnotesize Table~\thetable. #1}
        \ifdim \wd\@tempboxa > 5in
           {\begin{center}
        \parbox{5in}{\footnotesize\smalllineskip Table~\thetable. #1}
            \end{center}}
        \else
             {\begin{center}
             {\footnotesize Table~\thetable. #1}
              \end{center}}
        \fi}

\def\@citex[#1]#2{\if@filesw\immediate\write\@auxout
	{\string\citation{#2}}\fi
\def\@citea{}\@cite{\@for\@citeb:=#2\do
	{\@citea\def\@citea{,}\@ifundefined
	{b@\@citeb}{{\bf ?}\@warning
	{Citation `\@citeb' on page \thepage \space undefined}}
	{\csname b@\@citeb\endcsname}}}{#1}}

\newif\if@cghi
\def\cite{\@cghitrue\@ifnextchar [{\@tempswatrue
	\@citex}{\@tempswafalse\@citex[]}}
\def\citelow{\@cghifalse\@ifnextchar [{\@tempswatrue
	\@citex}{\@tempswafalse\@citex[]}}
\def\@cite#1#2{{$\null^{#1}$\if@tempswa\typeout
	{IJCGA warning: optional citation argument 
	ignored: `#2'} \fi}}

\def\pmb#1{\setbox0=\hbox{#1}
	\kern-.025em\copy0\kern-\wd0
	\kern.05em\copy0\kern-\wd0
	\kern-.025em\raise.0433em\box0}

\def\fnt#1#2{\footnotetext{\kern-.3em
	{$^{\mbox{\scriptsize #1}}$}{#2}}}

 \def\fpage1{\begingroup
 \voffset=.3in
 \thispagestyle{empty}\begin{table}[b]\centerline{\footnotesize 1}
	\end{table}\endgroup}

\def\ps@myheadings{%
   \let\@oddfoot\@empty\let\@evenfoot\@empty
    \def\@evenhead{\slshape\leftmark\hfil}
    \def\@oddhead{\hfil{\slshape\rightmark}}
    \let\@mkboth\@gobbletwo
    \let\sectionmark\@gobble
    \let\subsectionmark\@gobble
    }
\font\tenrm=cmr10
\font\tenit=cmti10 
\font\tenbf=cmbx10
\font\bfit=cmbxti10 at 10pt
\font\ninerm=cmr9

\font\eightrm=cmr8

\def\qed{\hbox{${\vcenter{\vbox{			
   \hrule height 0.4pt\hbox{\vrule width 0.4pt height 6pt
   \kern5pt\vrule width 0.4pt}\hrule height 0.4pt}}}$}}

\begin{document}
\setlength{\textheight}{7.7truein}  

\thispagestyle{empty}

\markboth{\protect{\footnotesize\it 
J. D. Anderson, E. L. Lau,  S. G. Turyshev, P. A. Laing, \& M. M. Nieto}}
{\protect{\footnotesize\it Search for a Standard
Explanation of the Pioneer Anomaly}}

\normalsize\textlineskip

\setcounter{page}{1}

\copyrightheading{}

\vspace*{0.88truein}


\centerline{\bf SEARCH FOR A STANDARD EXPLANATION}
\baselineskip=13pt
\centerline{\bf OF THE PIONEER ANOMALY}
\vspace*{0.4truein}
\centerline{\footnotesize JOHN  D. ANDERSON$^a$, EUNICE L. LAU$^b$, 
and  SLAVA G. TURYSHEV$^c$}

\centerline{\footnotesize\it 
Jet Propulsion Laboratory, California Institute
of Technology, Pasadena, CA 91109 U.S.A.}
\baselineskip=10pt
\centerline{\footnotesize\it  $^a$john.d.anderson@jpl.nasa.gov}
\centerline{\footnotesize\it $^b$Eunice.L.Lau@jpl.nasa.gov}
\centerline{\footnotesize\it V$^c$turyshev@jpl.nasa.gov}

\baselineskip=25pt
\centerline{\footnotesize  PHILIP A. LAING$^d$} 
\baselineskip=13pt
\centerline{\footnotesize\it 
The Aerospace Corporation, 2350 E. El Segundo Blvd.,El Segundo,
CA 90245-4691 U.S.A.} 
\baselineskip=10pt 
\centerline{\footnotesize\it $^d$Philip.A.Laing@aero.org}

\baselineskip=25pt
\centerline{\footnotesize  MICHAEL MARTIN NIETO$^e$}
\baselineskip=13pt
\centerline{\footnotesize\it 
Theoretical Division (MS-B285), Los Alamos National Laboratory,}
\baselineskip=10pt
\centerline{\footnotesize\it
University of California, Los Alamos, NM 87545 U.S.A.}
\centerline{\footnotesize\it $^e$mmn@lanl.gov}

\baselineskip=10pt

\vspace*{0.228truein}

\begin{center}
\begin{small}
~
\end{small}
\end{center}

\vspace*{0.23truein}
\abstracts{
The data from  Pioneer 10 and 11 shows an anomalous, constant, 
Doppler frequency drift that can be interpreted as an 
acceleration directed towards the Sun of 
$a_P = (8.74 \pm 1.33) \times 10^{-8} ~~{\rm cm/s}^2$.
Although one can consider a new physical origin for the anomaly, 
one first must investigate the contributions of the prime candidates, 
which are systematics generated on board.  Here we expand upon
previous analyses  of thermal systematics.  We demonstrate that 
thermal models put forth so far are not supported by the analyzed 
data.  Possible 
ways to further investigate the nature of the anomaly are proposed. 
}{}{}


\vspace*{2pt}

\baselineskip=13pt	        
\normalsize              	


\section{Introduction}
\vspace*{-0.5pt}
\noindent
We have reported an anomalous, constant acceleration of Pioneer 10
and 11, $ a_P = (8.74 \pm 1.33) \times 10^{-8}$  cm/s$^2$, directed
towards the Sun, at  distances $\sim 20-70$ AU.\cite{usnew,anderson} 
The veracity of the signal is now  
undisputed, although the source of the anomaly, some systematic or
some not understood physics, is subject to debate.  

Since the announcement,\cite{anderson} attention focused on two main 
possible on-board systematic explanations for the anomaly.  The first 
is gas leaks, whose proponents tend to come from the space navigational
community.  In our detailed paper \cite{usnew} we gave a
thorough discussion on the limitations of this mechanism.  

The other most commonly discussed systematic is one due to heat,
either direct radiation from the RTGs (radioisotope thermoelectric
generators) or heat radiated from the central bus where most of the
electrical power is dissipated.   This mechanism tends to be favored by
space physicists.  Previous proposals for both of these sub-mechanisms
have been made.\cite{katz,murphy}  It was found that the data
and design characteristics of the craft appear to rule them 
out.\cite{uskatz,usmurphy,usnew}  

Nevertheless, discussion has continued.  Here we investigate other, more
complicated, heat mechanisms.  We find that, although the data is not
completely constraining, it does  not provide
evidence for these mechanisms.   
We conclude with ideas on how future work could, in principle, allow
one either to rule out an on-board systematic origin to the anomaly, 
meaning the explanation could be unknown physics,  
or to determine directly that some on-board systematic is 
the origin of the anomaly.


\setcounter{page}{2}

\section{Constant Electrical Heat Radiation as the Source}
\noindent
It has been recently suggested that most, if not all, of the unmodeled
acceleration  of  Pioneer 10 and 11 is due to an  essentially constant
supply of heat coming from the central compartment, directed out the
front of the  craft through the closed louvers.\cite{scheffer}  This
is a more subtle version of an earlier proposal\cite{murphy} 
calling on the total electrical power as a mechanism. 
That proposal was argued against because of the observed lack of
decay  of the acceleration with time.\cite{usmurphy,usnew}   

But this first hypothesis does not work.   The Pioneer spacecraft  were not
built and do not work that way.   (It was known beforehand that during the
extended mission there would be  ``dissipation of 70 to 120 W of
electrical power by units within the compartment.''\cite{dissipateref})
The assumption of constancy is incorrect for two reasons.

One reason for the incorrectness of this proposal\cite{scheffer} 
is that  the ``central compartment'' consists not just
of the hexagonal  ``equipment  compartment'' but also of the ``experiment
compartment.''  These two components are openly (radiatively) connected,
separated on their common side by only a half-plate wall, 
``partial vertical plate.''\cite{piodoc}   
(If one wanted to one could open the top of the
equipment  compartment and stick one's arm into the experiment
compartment.)  Indeed, to help with heat dissipation from the experiment
compartment  during the early stages of the flight,  there are louvers
placed on the front of the experiment compartment that are similar to
those placed on the equipment compartment.

Another reason why a proposal of constant heat as 
a source is incorrect is that
even the heat dissipated only from the ``equipment compartment'' is not  a
constant.  The lack of constancy of heat dissipated by itself  invalidates
the  hypothesis.  This is even without showing that correctly calculating
the insulation/louver properties also rules out the hypothesis.


\section{Detailed Properties of the Craft and the Data}
\noindent
Although the hypothesis of constant heat as a source is shown  
in the above Section to be ruled out, 
it is useful to demonstrate that the spacecraft design and data 
give added support to this ruling.  These details are presented in this
Section.   The details will also
be useful for the discussion of a second proposal in Section 4.  


\subsection{Electrical power and the louvers}
\begin{figure}[h]
\begin{center}
\indent
      \psfig{file=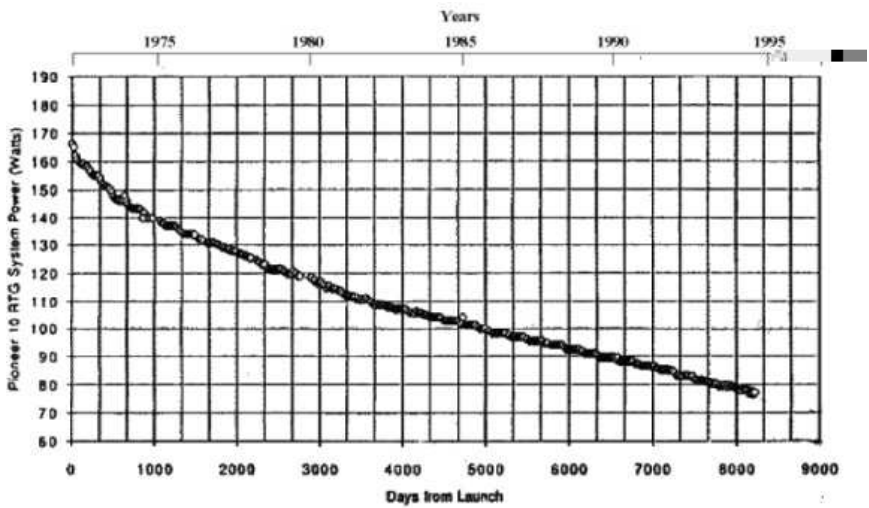,width=4.5in}
\end{center}
\caption{
The Pioneer 10 electrical power generated at the RTGs as a function of
time from launch on 22 March 1972, to near the end of 1994.   
By 1998.5, only $\sim$68 W was generated.
\label{epower}}
\end{figure}

\noindent
We begin by reviewing the  total
electrical power of the craft  {\it generated at the RTGs}.  (See Figure
\ref{epower}.)

After launch, when there quickly became a stable $\sim 165$ W of power, the
electrical power for Pioneer 10 first decayed at a rate of $\sim 10.6$ W/yr.  
After Jupiter encounter the power decay rate changed to a new value of
$\sim 4.4$ W/yr.  Finally, after around 1987 the decay rate changed to
$\sim 2.6$ W/yr.  In particular,   
at the beginning of our run (1987.0)  $\sim 97$ W was generated 
and at the end of our run (1998.5)  $\sim 68$ W was 
generated.\cite{usnew}  The power
decay rate (which was higher for Pioneer 11\cite{pio11}) was caused by 
degradation of the thermistor junctions and other RTG electrical
components on top of the smaller radioactive decay rate of the 
$^238$Pu thermal sources (a half-life of 87.74 yr or a decrease of $\sim$0.8\% 
each year). 

What happens to this available power?  We normalize to the end of our
run,  1998.5.   About 3 W goes into external IR cable losses.  The 
remaining 65 W power enters the central instrument compartment.  There 
it is approximately  used as power for the Inverter Assemblies (7 W), 
the Central Transformer-Rectifier-Filter 
[CTRF] and subsystems (21.1 W), a separate TRF (2 W)
the Traveling Wave Tube Amplifier [TWTA] (27.8 W), and 
the Power Control Unit [PCU] and battery (3 W). 
Other small electronic components and base shunt-current loss account for a
few W.  Therefore, at this late date there are at best only a few W
left for the experiments or anything else.  
For example, in  a  recent July 2000 maneuver, which 
required enough power to run the attitude and conscan subsystems, 
the TWTA had to be turned off, ending its continuous activation.
It was turned back on after the 2000 conscan and then off and on for
the 2001 and 2 March 2002 conscans.

Next, as can be verified in any of a number of 
references,\cite{piodoc,extended,piopr2} there  are 
8 W $= 10\log_{10}[8000]$ dBm = 39 dBm of constant radio power being 
emitted in a collimated beam towards the Sun.   
[Actually, the beam is along the spacecraft spin axis which,
during the time of our data set (1987-1998.5), on average points
towards the Sun.]  This is presently our largest systematic which 
works against $a_P$ and hence makes the final value {\it
larger} than the experimental number.   This means the total possible heat 
power for the spacecraft bus is about 57 W, no matter how it escapes.    
(And remember, a constant $\sim$ 63 W of totally directed power are 
needed to explain $a_P$.\cite{usnew})  

\begin{figure}[h]  
\begin{center}
\indent
    \psfig{file=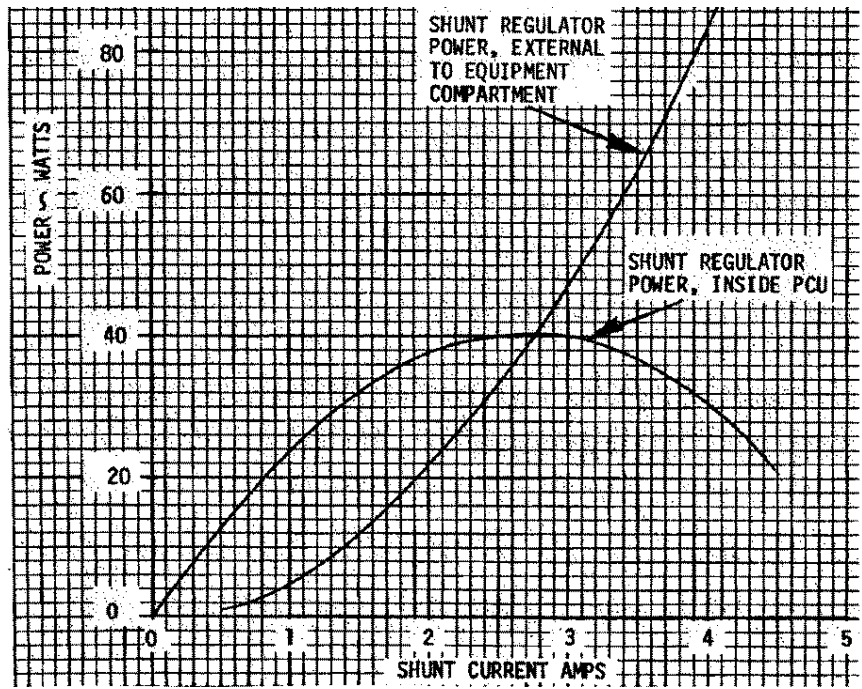,width=4.5in}
\end{center}
\caption{Design shunt power distribution as a function of 
shunt current.  
\label{shuntcurrent}}
\end{figure}

Now, recall that the louvers were open during the early mission to
let heat escape more easily since there was much more excess
electrically-generated heat.  
Further,  in the early stages
there was also much more heat to be dissipated through the 
shunt external regulator (called radiator) and internal regulator.    
When there is high shunt current, the majority of the dissipation is 
directed to the {\it exterior} shunt radiator. 
(See Figure \ref{shuntcurrent}.\cite{piodoc}) 
These conditions were to  keep the temperature 
within the central compartment from being too high.  
Later, when the louvers were closed,\cite{usmurphy} the problem was the 
opposite, to try to keep
in as much heat as possible in order to maintain 
the constantly falling temperature. In this lower-power situation, the
majority of the (smaller) shunt dissipation is interior to the compartment.  
Further, the louvers when closed are designed not to radiate heat but to 
{\it retain} heat.  There are second surface mirrors on the insides of 
the louvers.  (For more information see our reference.\cite{heathold})

This indicates that the heat from the compartment is not radiated in a
strongly directional sense along the spin axis.  
But this appears to be in conflict with 
the values for the effective emissivities 
of the insulation and louver outside material (0.01 and 0.27,
respectively) that were used in our reference.\cite{scheffer}

How can that be?\cite{cassini}
Because the assumptions are off by large amounts (in 
opposite directions) from the real emittances  ($\sim$ 0.70 and 0.04, 
respectively).\cite{piodoc,em}  In fact saying all the 57 W goes out the 
louvers with an area of $\sim$1 m$^2$ and 
using the Stefan  Boltzmann law would
mean the exterior louver temperature was 398 K.  
This explains why it is difficult to understand the proposed 
through-the-louvers, directed-heat, emission  mechanism.\cite{cassini} 


\subsection{Electrical power dissipation history}
\noindent
Further,  and as we now come to, 
the observed approximate constancy of the anomaly over the 
data period\cite{constant} is in conflict with any model that 
says the anomaly is from a  constant compartment source.\cite{scheffer} 
The heat in the compartment is {\it not} constant.

Consider the situation at the beginning of our analyzed data.  In 1987.0,  
there was about 97 W  total electrical power or about 
29 W more than at the end of our run.  Where did this go?  
Approximately 1 W more went to cable loss,   
about 5 W more  went to higher Inverter Assembly/TRF losses, and 
$\sim$24 W went to run all the instruments.  Of this at least 
12 W went into the equipment compartment because 
the instruments are there.   (Even a few of the
external experiments also have their electronics there.)  
Further, the external Asteroid/Meteoroid detector (1.7 W) failed at 
Jupiter encounter and the external magnetometer (HVM) (3 W)
failed just before our data set, so
this power also would be going to the interior shunt regulator. 
[We note again that for low shunt current almost all of the heat loss 
was designed to be through the internal shunt regulator.  See Figure
\ref{shuntcurrent}.]  Therefore, at the beginning of our data set  
the size of the effect of this thermal hypothesis should be on the 
order of $(73/57)= 1.26$ times the size of effect at the end of our run.  

\indent From the beginning to the end of our data,
the ODP \cite{odp} results (before systematic biases are included) 
differed by only 0.19 U out of a total final effect of 8.74 U.   
[We define $1~U \equiv  1 \times 10^{-8}$ cm/s$^2$.]
We took this difference to be due to spin-rate change.  But even if
it is not (and granted uncertainties) there obviously is no measured 
difference on the order of 25\%.  (Observe that any RTG radiant-heat based
systematic\cite{katz,uskatz} must also decay by 8\% during our data run.)

Going further back might allow us to get an even better handle on this. 
In Figure 7 of reference\cite{usnew} we showed preliminary ODP results 
for Pioneer10 using data from approximately  1981.5 to 1989.4.  These 
results were not spin-rate change adjusted,\cite{usnew} were not treated 
for systematics,  
used different time-evolving estimation procedures,\cite{usnew} 
were done by three separate JPL navigation specialists, 
separated and smoothed by one of us,\cite{JDA} and definitely 
not analyzed with the care of our recent run (1987.0 to 1998.5).
The value of $a_P$ at the end of this earlier run is similar 
to the value at the beginning of our present data run,\cite{usnew} 
which overlaps it in time.  

At the beginning of this earlier data set the total
electrical power was $\sim$116 W.  Where did this added  19 W go? 
About $\sim$1 W more went into cable losses.  Of the remaining 18 W, 
it is expected that $\sim$5 W went into added internal 
Inverter Assembly/TRF losses, $\sim$11 W 
into the internal shunt regulator, and only of order $\sim$2 W
into the external shunt radiator. Therefore, in 1981.5 this thermal  
hypothesis should have produced an effect $\sim (89/57) = 1.56$ times 
that at the end of our run in 1998.5.  However, 
there is no fractional change in size of $a_P$ to indicate that 
most if not all of the anomalous acceleration is due to
electric-power heat.  (Also, during this 
total period any RTG radiant-heat based systematic\cite{katz,uskatz}
would have decreased by 13\%.)

\begin{figure}[h]
\begin{center}
\indent  
    \psfig{file=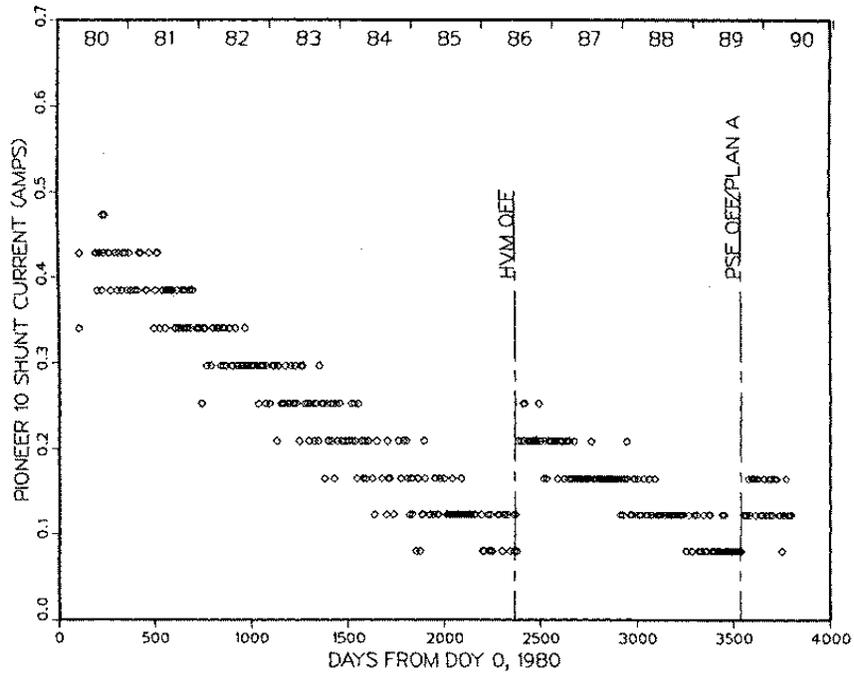,width=4.5in}
\end{center}
    \caption
{The Pioneer 10 shunt-current history from 1980.0 to 1991.0.
(The data is in bimodal form, with steps of $\sim0.05$ Amps.)
      \label{shunthistory}} 
\end{figure}

One can verify these conclusions by consulting
the shunt-current history.  After our run started, there was basically no 
excess power to get rid of unless instruments were turned off and on
(power-sharing).  After 1992.0 the shunt-current history shows a 
current between 0.09 and 0.14 Amps with various spikes.  
(0.9 W is the base internal shunt power loss.)  
Before 1990.5 the situation is approximately as 
we have described. (See Figure  \ref{shunthistory}.) 

One-half year before the start (1987.0) of our data run, the external
HVM shut off.  This mainly external heat of 3 W then went 
into internal shunt heat via an additional 0.15 Amps of shunt
current.  In  our data run, one can first see the shunt current 
decaying, and then the  PSE (Program Storage and Execution Unit) was
turned off.  Before 1987.0 the shunt current is increasing.
So, the agreement with our description is clear.   


\section{Incorrect Solar Radiation Dynamics as an Added Source} 
\label{rheat}
\noindent
To correct the deficiencies of a constant heat source as the origin of
the anomaly, it was later proposed  that the spacecraft solar 
radiation constant is off by a factor +0.3.\cite{schefferB}   
Then it was argued that this new contribution,
plus the contribution from the model of the last section, and perhaps
RTG radiation could all add up to produce 
the anomaly.  But if this new contribution (from incorrect
solar-radiation modeling of the craft) were true the Pioneer Mission
would have failed!

In the first place,  the true solar radiation model is much more
complicated than the simple  model that is 
proposed.\cite{schefferB}  The true model involves many
parameters describing physical quantities of different parts of the craft
and the angle of orientation with respect to the Sun.  This was {\it very} 
important during the early phase of the mission.  In fact,  
by the second day of the mission   
the Pioneer 10 aim point was already 
several Jovian radii off and on the wrong side of Jupiter.   
To correct these navigation errors, 
a newly developed, complicated, solar radiation model\cite{rad} 
(the HGA alone being like a solar sail)   
was implemented in POEAS (a progenitor of CHASMP).\cite{laing}    
As a result, a course correction was implemented, and Pioneer 10 
successfully reached its targeted aim point at Jupiter.  
The model was then utilized in ODP, confirming the solution.

The next time a solution for the 
solar radiation model parameters was required was when the
Pioneer 10 spacecraft was still in the 
Jovian vicinity, after the successful fly-by. That solution was
obtained without any complications using the same model. The
solution was also stable, reliable, and within acceptable 
limits.\cite{usnew}  
Finally, a later fit to near-Earth Pioneer 10/11 data 
verified all the model parameters to $\pm 5$\%.\cite{null76}  
This error is much smaller than the 
change of the parameters assumed in the ``incorrect radiation 
dynamics'' explanation of the Pioneer anomaly.\cite{schefferB} 

Even further, and as we already have noted, 
the early numbers in Figure 7 of reference\cite{usnew} were based on 
attempts to check the orbit accuracy.\cite{JDA}  In particular, the 
first two Pioneer 11 points, included in the early memos,\cite{JDA} were at
the distances of Jupiter and Saturn encounters.  Many things, such as
large maneuvers were going on; Pioneer 11 encountered Jupiter
and then came back across the central solar system to encounter Saturn. 

One must be careful if one assumes a measurement is wrong just to obtain the
theoretical systematic result one expects.  One should try to apply the same
standards in obtaining a ``normal explanation'' as one would if one
were proposing a radical one. 


\section{Possible Future Work} 
\noindent
Of course, at present our arguments are good to only a few W.  
We might be able to improve upon our knowledge by studying in detail the 
instrument/power usage history.  
But this could be of no use because the data may be lost or
inadequate.  However, there is 
more one can do to provide an even more precise systematic.  After all,
heat in some form remains a primary candidate 
(perhaps the best candidate) for an explanation from 
onboard-generated  forces.  It is
just that so far no explicit model has been shown to work. 

The latter part of the preliminary (between 1981.5 and 1989.4) data  
overlaps the beginning of our later (between 1987.0 and 1998.5) 
analyzed data set.\cite{usnew}  Therefore,  we can normalize 
the preliminary data to our reported\cite{usnew} 
anomalous ``experimental'' value of 7.85 U, which was corrected
for spin-rate change but not for other systematics.  The limited 
spin-rate data we have available in our Pioneer data archives 
shows no dramatic or anomalous spin-rate changes  
before 1989 and definitely no anomalous long-term change over 1981.5
to 1987.5 (after which our archiving of all spin data is complete).  

Does this therefore mean, then, that in Figure 7 of reference 
\cite{usnew} we see a shift in  $a_P$ of $\sim$0.6 U as we go 
back to 1981.5?   It is not clear.  A shift of $\sim$0.6 U 
can not be determined on the basis of only this simple argument. 
The situation is much too complicated.

In addition to the problems mentioned at the 
end of the previous Section, the analyses used in the preliminary
plot were done before an annual sinusoid was characterized.  
This annual term,  which can be removed by an unreasonably large
adjustment to the Earth's orbital orientation (inclination and node on
the J2000 Earth equator), is larger closer in to the Sun.\cite{usnew} 
All of the high data points on the preliminary $a_P$ 
curve are at boundaries of years, when the annual term is at a maximum.  
So, for example, the hypothesis discussed in Sec. 4  
would imply that step-function 
changes in the heat being dissipated internally 
(see Figure \ref{shunthistory}) would make the sensitive annual term 
show abrupt deviations from its sinusoid. 
This is not seen in our primary data run. 

The use of the shunt-current history and temperature 
history data would allow correlations of any $a_P$ variation with 
the compartment power-sharing.  
when  (i) all systems and experiments were on with a significant 
shunt current, and (ii) the instruments were being turned off and
there was little shunt current.  

An analysis to improve the characterization of $a_P$
from the earlier data, although desirable, would be difficult. 
The earlier radio Doppler and spin-rate data is available in a 
not completely processed format.\cite{eunice}  (They exist on 
obsolete 9-track magnetic tapes intended for a {\small VAX}.)  

A precise analysis over the entire 17 years of available 
Pioneer data (perhaps also including later data) might 
reveal any changes in $a_P$ that vary slowly  with time.
Any changes could be correlated with the radioactive decay rate of the 
plutonium and the electrical power history.   
Then a precise heat systematic based on data rather than on
`back-of-the-envelope' thermal models might be found.  

Further, it might be possible to do a statistically significant 
three-dimensional acceleration analysis,\cite{3D} 
especially with the early data near 1981.    
If so, one could distinguish an anomalous acceleration towards the Sun
from an acceleration along the spin axis.  

Finally, a well-designed new craft and experiment could be undertaken
to take advantage of what we have learned.\cite{mission}  This
basically would be a new gravitational mission with the goal of
unambiguously measuring the Pioneer anomaly to an accuracy of 10\% or better.


\section{Conclusion}
\noindent
In conclusion, we quote from ourselves.\cite{usnew} ``Until more is known,
we must admit that the most likely cause of this 
effect is an  unknown systematic.''  
As for heat,  there is no solid explanation in hand
as to how a specific mechanism could work.  Most importantly,  the 
decrease in the heat supply over time should have been seen by now. 
It is not.  To further quote from ourselves,\cite{usnew}  
``... we anticipate that, given our analysis of the Pioneers, in the 
future precision orbital analysis may concentrate more on systematics.''  
This may indeed be the most important outcome of our analysis,   
with important implications for future deep-space missions, including 
the hoped for Pluto/Kuiper Belt mission.\cite{lisa} 


\nonumsection{Acknowledgements}
\noindent
Over the past four years we have received many communications on the
Pioneer spacecraft and their thermal properties, from numerous
individuals.  Among them have been D. F. Bartlett, E. Batka, R. Jackson,
J. I. Katz, L. Kellogg, L. Lasher, D. Lozier, E. M. Murphy, R. Ryan, L.
K. Scheffer, and V. J. Slabinski.  

This work was sponsored by the 
Pioneer Project, NASA/Ames Research Center, and was performed at the Jet
Propulsion Laboratory, California Institute of Technology, under contract
with the  National Aeronautics and Space Administration.
P.A.L.  was supported by a grant from NASA through the
Ultraviolet, Visible, and  Gravitational Astrophysics Program. M.M.N.
acknowledges support  by the U.S. DOE.


\vspace*{2pt}
\nonumsection{References}


\begin{thebibliography}{000}
\bibitem{pio11} If this type of mechanism were the main 
source of $a_P$ then, since Pioneer 11 radiated less electrical 
heat in the extended mission, 
$a_P$ for Pioneer 11 should  be less than that for Pioneer 10. 
Experimentally it is the opposite.
\bibitem{constant} The 1-day
batch-sequential result is discussed in our reference.\cite{usnew}
\bibitem{eunice}  The ATDF's  (Archival Tracking Data Files) were 
processed through the STRIPPER software for input into an orbital 
data program.  (See our reference.\cite{usnew})  The tracking data tapes 
themselves are in binary form in a more basic format.  The observable
and residual has not been calculated in the form which is used by the ODP.
\bibitem{lisa} It was suggested\cite{schefferB} one could use LISA
to study the Pioneer anomaly.  
However, the three spacecraft of the LISA constellation 
will be on  bound, interior elliptic orbits  close to 1 AU 
whereas the Pioneers are on unbound, hyperbolic 
orbits beyond 20 AU.
\end{thebibliography}
\end{document}